\documentclass[prl,aps,twocolumn,floats,floatfix,showpacs]{revtex4}
\usepackage{graphicx}
\usepackage{times}

\newcommand{\be}{\begin{equation}}
\newcommand{\ee}{\end{equation}}
\newcommand{\bea}{\begin{eqnarray}}
\newcommand{\eea}{\end{eqnarray}}
\newcommand{\la}{\left<}
\newcommand{\ra}{\right>}
\newcommand{\lb}{\left[}
\newcommand{\rb}{\right]}
\newcommand{\lp}{\left(}
\newcommand{\rp}{\right)}
\newcommand{\lf}{\left\{}
\newcommand{\rf}{\right\}}

\newcommand{\p}{{\bf p}}
\renewcommand{\r}{{\bf r}}
\renewcommand{\u}{{\bf u}}

\newcommand{\K}{{\cal K}}

\newcommand{\tee}{\tau_{\rm qp}}
\newcommand{\dn}{\delta \tilde n}
\newcommand{\f}{{\sf f}}

\newcommand{\EF}{{\epsilon_{F}}}

\begin{document}
\title{Viscosity of a two-dimensional Fermi liquid}
\author{D. S. Novikov}
\email{dima@alum.mit.edu}
\affiliation{Department of Electrical Engineering and Department of Physics, 
Princeton University, Princeton, New Jersey 08544}
\date{\today}

\begin{abstract}

Shear viscosity of a two-dimensional Fermi liquid is found to be a 
nonanalytic function of temperature.
In contrast to the quasiparticle lifetime that is determined by the 
forward-scattering processes, 
the main contribution to the viscosity arises from the quasiparticle scattering
in the Cooper channel. 
The viscosity is enhanced by the logarithmic singularity of the vertex part.
This singular behavior can manifest itself 
in the two-dimensional electron transport,
and in the momentum relaxation of fermions in atomic traps.

\end{abstract}


\maketitle

Viscosity characterizes momentum relaxation in a fluid. 
In the presence of a weakly inhomogeneous flow $\u(\r)$, the linear relation
between the stress tensor of the $d$-dimensional fluid 
and the flow gradients is conventionally written as \cite{Landau10}
\be \label{def-eta}
\Pi_{ij} = - \eta 
\lp \partial_i u_j + \partial_j u_i -\frac2d \delta_{ij} \nabla \u \rp 
- \zeta \nabla \u \,.
\ee
The coefficients $\eta$ and $\zeta$ are the first (shear) and the second (bulk)
viscosities, out of which $\eta$ usually dominates: $\eta \gg \zeta$. 
The linear relation (\ref{def-eta}), complemented by the 
momentum conservation law 
$\rho \lp \partial_t + \u\nabla\rp u_j = -\partial_i \Pi_{ij} \simeq 
\eta \nabla^2 u_j$,
where $\rho(\r)=mN(\r)$ is the fluid mass density,
leads to the Navier-Stokes description on the length 
scales exceeding the molecular mean free path $\ell$.
Such a hydrodynamic approach bears great predictive power, since 
a single parameter $\eta$ accounts for the details 
of the molecular interactions on the scale $\ell$. 
The dependence of the viscosity on the temperature and density  
directly determines the fluid's relaxation and transport properties.
Calculating $\eta(T,N)$ is in general a difficult task.

Early developments of the kinetic theory by Chapman and Enskog 
have lead to the understanding of the viscosity of dilute gases. 
The latter can be estimated as \cite{Landau10,Smith-Jensen-book}
\be \label{eta=vl}
\eta \sim \rho \bar v \ell \,, 
\ee
where $\bar v$ is the typical velocity of a molecule in a gas, and 
the mean free path $\ell$ is determined by the molecular density 
and the scattering cross-section. Assuming weak energy dependence of the
scattering, one readily obtains the universal temperature dependence 
of the gas viscosity $\eta \propto \sqrt{T}$ that 
stems from that of $\bar v$ according to the Maxwell distribution.

Later on, the notion of viscosity was generalized 
for hydrodynamic modes in quantum fluids \cite{Abrikosov-Khalatnikov}.
In particular, the shear viscosity 
has been studied in the context of 3d Fermi liquids (FL), 
with applications to liquid $^3$He.
Remarkably, the simple estimate (\ref{eta=vl}) qualitatively applies in
the degenerate case, since at $T\ll \EF$, where $\EF$ is the Fermi energy,
the FL is a dilute gas of quasiparticles \cite{AGD}. 
Moreover, 
Eq.~(\ref{eta=vl}) yields a singular temperature dependence $\eta \propto 1/T^2$.
The latter originates entirely from that 
of the quasiparticle inelastic mean free path $\ell = v_F \tee(T)$, with 
$\hbar/\tee \propto T^2/\EF$  
the scattering rate at the Fermi surface in 3d \cite{T2,AGD}, while 
$\bar v = v_F$ is $T$-independent. Physically, the $T\to 0$ divergence signifies 
the increasing resistance to the shear flow $\u(\r)$ that causes
distortions of the Fermi surface.

The characteristic dependence $\eta \propto 1/T^2$, 
first estimated by Pomeranchuk 
in 1950 \cite{Pomeranchuk'1950,Landau10} on dimensional grounds,
was later confirmed by the calculations based on the FL kinetic equation 
\cite{Abrikosov-Khalatnikov,Brooker-Sykes,Smith-Jensen-book}.  
It was also shown that the second viscosity $\zeta$ is 
practically irrelevant, $\zeta \sim (T/\EF)^2 \eta$ \cite{Abrikosov-Khalatnikov}.
%
Subsequent $^3$He measurements \cite{visc-He3}
confirmed the relation $\eta T^2 = {\rm const}$.

In the present work we consider 
the shear viscosity of the two-dimensional Fermi liquid. 
From a practical standpoint, the problem is relevant to a variety of Fermi systems,
ranging from the 2d electrons in heterostructures, to the trapped
Fermi gases. Theoretically, the problem is compelling since in a 2d FL one 
generally expects nonanalytic energy- and temperature-dependence of response 
functions due to strong restrictions on the quasiparticle scattering 
\cite{2d-nonanalyt}.
Indeed, a naive estimate of the kind (\ref{eta=vl}) 
should give a logarithmic suppression 
$\eta \sim \tee \sim 1/T^2\ln (\EF/T)$ due to the well-known 
enhancement of the quasiparticle scattering rate
$\tee^{-1} \propto (T^2/\EF) \ln (\EF/T)$ 
\cite{HSW,Chaplik,Bloom,GQ,FA,ZDS,JMD,ML,RW,NZA}.
Quite unexpectedly, we find that the 2d viscosity is {\it enhanced} by the 
square of the large logarithm $\ln (\EF/T)$, as compared to that of the 3d FL:
\be \label{eta-2d}
\eta \simeq {3\over \pi} {N \hbar\over F_\pi^2} 
\lp {\EF \over T}\rp^2 \ln^2\lp {\EF \over T}\rp .
\ee
The behavior (\ref{eta-2d}) originates from 
the Cooper-channel processes in which the sum ${\bf s}=\p_1+\p_2$ 
of the colliding momenta is much smaller than the Fermi momentum, $s\ll p_F$.
The logarithm arises from the corresponding singularity of the vertex part 
$\Gamma \propto F_\pi/\ln(p_F/s)$ \cite{AGD,AleinerEfetov},
while the quasiparticle collisions at other angles appear to be less relevant.

The temperature dependence (\ref{eta-2d}) 
can manifest itself in the form of the interaction correction to the 
transport properties of 2d electron systems in the presence of a smooth 
disorder potential. 
The temperature-dependent hydrodynamic contributions to conductivity, 
$\sigma \propto 1/\eta(T)$, caused by the resistance to the 
laminar flow of an electron liquid, 
have been anticipated since 1960's \cite{Gurzhi-60s}.
Recently the role of hydrodynamic modes in electron transport was 
revealed by observing the switching from Knudsen to Poiseuille flow of the
electrons in 2d wires \cite{Molenkamp,Gurzhi-95}. 
Assuming the laminar flow in the bulk 2d samples, the result (\ref{eta-2d}) 
suggests that in $d=2$ the hydrodynamic modes may lead to a singular temperature 
dependence of the resistivity $R_{xx}\propto \ln^2 T/T^2$. 
This prediction can be relevant for transport measurements
in clean 2d heterostructures in the metal-insulator transition regime: 
The interaction between the delocalized carriers may cause the 
apparently ``insulating'' correction to transport \cite{Spivak}.
Viscosity may also limit the functionality of the FET devices
suggested as a means of plasma waves generation \cite{Dyakonov-Shur}.
Finally, hydrodynamic modes can play a role in the dynamics 
of the trapped atomic Fermi gases \cite{traps-exper,Bruun-Smith}. 
The result (\ref{eta-2d}) points at a possibility of measuring the 
Cooper-channel amplitude $F_\pi$ (defined below) 
as a functional of the repulsive interaction between the fermions. 

In what follows, 
we consider the 2d FL Boltzmann equation, and draw a distinction
 between the scattering contributions to the 
quasiparticle lifetime and to the viscosity.

{\it The Boltzmann equation.---} The kinetic equation for the FL 
\be \label{BE}
\partial_t n + \partial_{\r} n \partial_{\p} \epsilon 
- \partial_{\p} n \partial_{\r} \epsilon = {\rm St}\lf n\rf , 
\ee
where the quasiparticle energy 
$\epsilon_\p = \epsilon^{(0)}_\p + \delta \epsilon$, 
$\delta \epsilon = \int \f_{\p\p'} \delta n_{\p'} d\tau_{\p'}$,
is the functional of the distribution function $n=n_0(\epsilon^{(0)}) + \delta n$,
and $\f$ is the Landau function [$d\tau_\p \equiv 2 d^2\p/(2\pi\hbar)^2$].
We assume the presence of a small nonuniform velocity flow $\u(\r)$
such that the equilibrium distribution function is 
\be \label{n0}
n_0 = f\lb (\epsilon_\p^{(0)} - \p \u - \EF)/T \rb , 
\quad f(x) = (e^x +1)^{-1} \,.
\ee
The linearized LHS of Eq.~(\ref{BE}) is obtained by 
setting $n\approx n_0$ and keeping the flow gradients $ \partial_i u_j$ in the 
second term: 
\be \label{LHS}
- {\partial n_0\over \partial \epsilon} 
\lp p_j {\partial \epsilon\over \partial p_i} - {\delta_{ij}\over d} 
\p {\partial \epsilon\over\partial \p}\rp \cdot \frac12 
\lp \partial_i u_j + \partial_j u_i - \delta_{ij} \nabla \u \rp \,.
\ee
The first and the third terms of Eq.~(\ref{BE}) almost cancel each other.
Their difference $\sim (T/\EF)^2$ yields
$\zeta\sim (T/\EF)^2 \eta$ \cite{Abrikosov-Khalatnikov}.

\begin{figure}[b]
\includegraphics[width=1.7in]{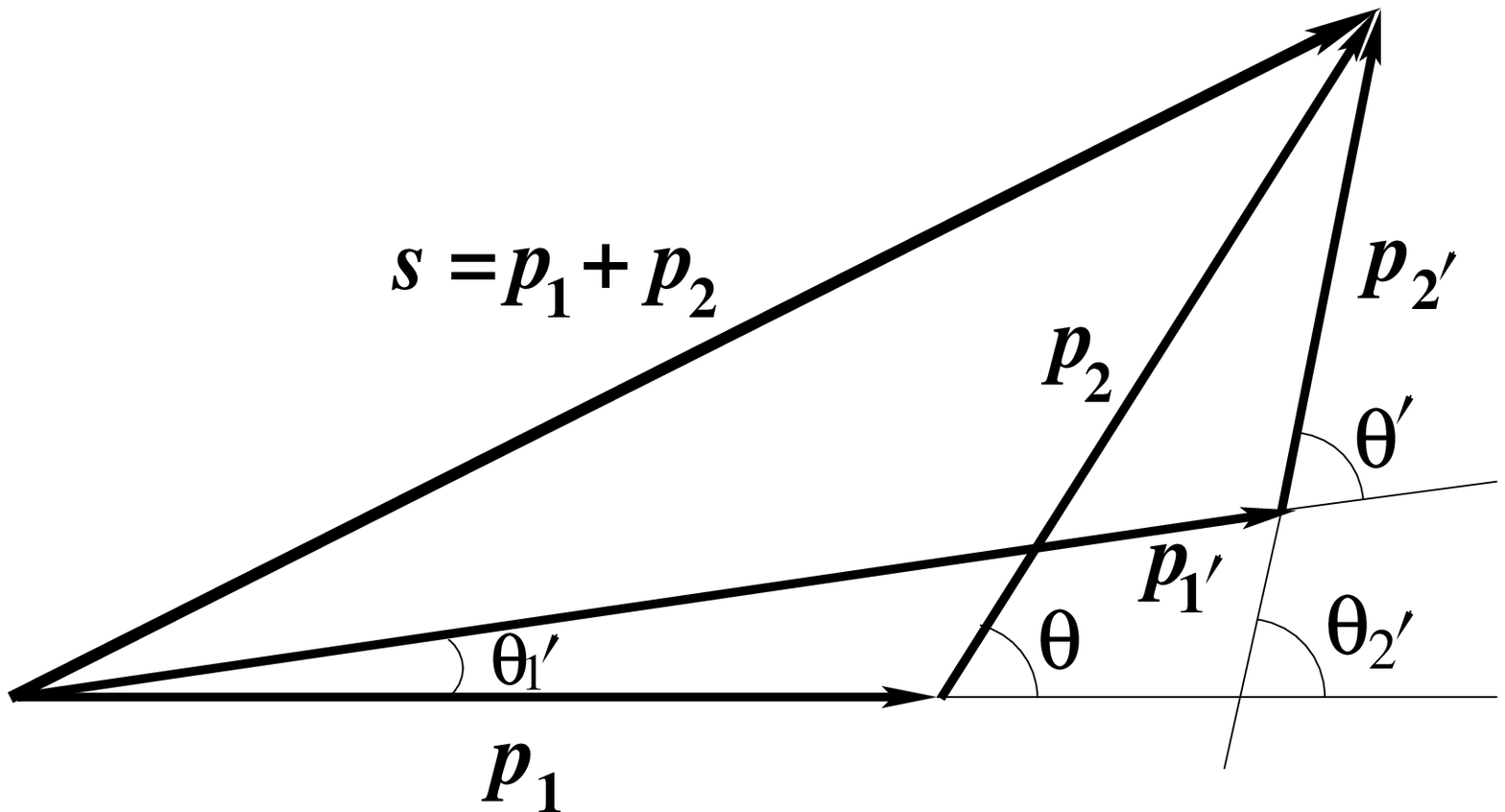} \ \ \
\includegraphics[width=1.5in]{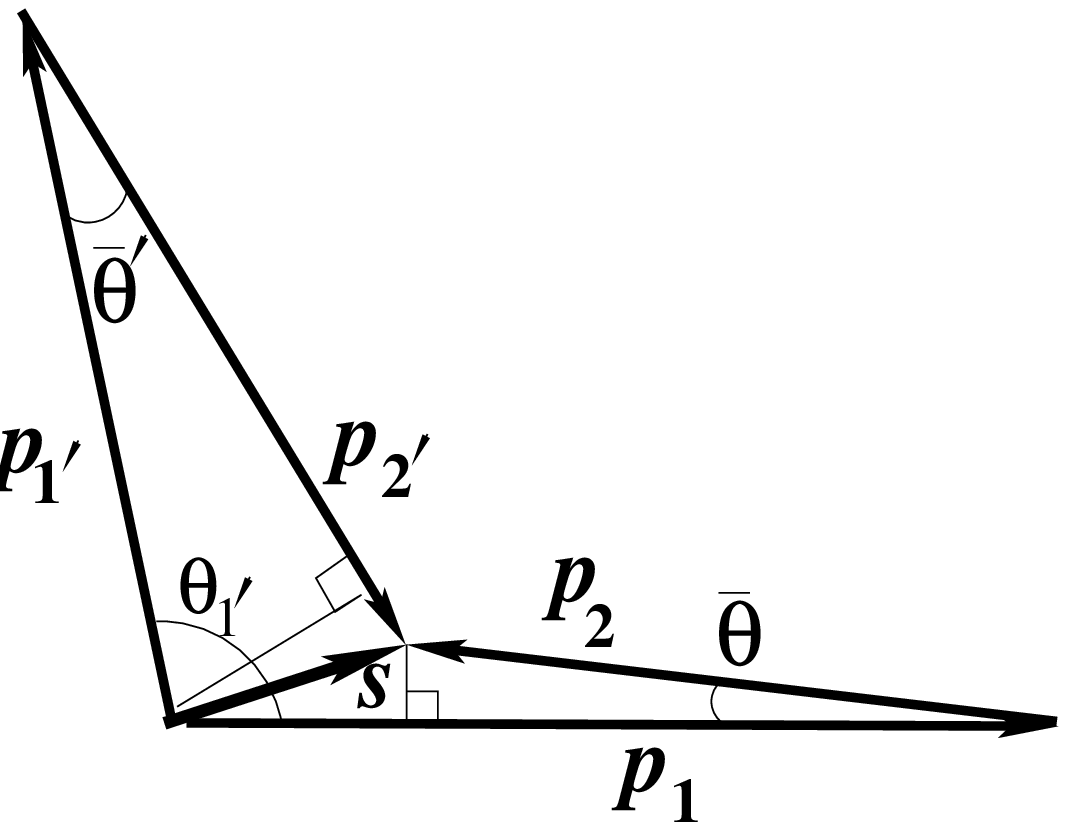}%
\caption[]{
Quasiparticle collisions on a plane.
At finite temperature $T\ll \EF$,
the incoming $\p_1, \ \p_2$, and outgoing $\p_{1'}, \ \p_{2'}$ momenta
can slightly differ in magnitude from $p_F$. 
Left: a generic collision. Right: the Cooper-channel process, 
$\bar\theta,\bar\theta'\ll 1$ and $\theta_{1'}\simeq\theta_{2'}\sim 1$.
}
\label{fig:collision}
\end{figure}

Quasiparticle collisions conserve energy and momentum.
The collision term ${\rm St}\lf n\rf$ vanishes 
for the equilibrium distribution $n_0(\epsilon)=n-\delta \tilde n$, where 
$\epsilon$ is the true quasiparticle energy, and the deviation 
$\delta \tilde n = \delta n - (\partial n_0/\partial\epsilon) \delta \epsilon$
includes the FL corrections. 
As usual, we identify the
smooth part of $\delta\tilde n$ as 
$\delta\tilde n = - (\partial n_0/ \partial \epsilon) \psi$,
where ${\partial n_0/ \partial \epsilon} = - {n_0 (1-n_0)/ T}$.
The resulting linearized collision term written in terms of $\psi$
has the form similar to that of a weakly-interacting Fermi gas:
\bea \nonumber
{\rm St} \lf n\rf = - \frac1T \int\! d\tau_2 d\tau_{1'} 
w(\theta) n_{01} n_{02}(1-n_{01'})(1-n_{02'}) \\
\times
\delta(\epsilon_1 + \epsilon_2 - \epsilon_{1'}-\epsilon_{2'})
(\psi_1 + \psi_2 - \psi_{1'}-\psi_{2'}) \,. \quad
\label{St-psi}
\eea
Here $\p_1 + \p_2 = \p_{1'}+\p_{2'}$ is assumed (see Fig.~\ref{fig:collision}),
$\theta=\angle(\p_1,\p_2)$, and the scattering probability
$w(\theta) \simeq (2\pi/\hbar) |z^2 \Gamma(\p_1,\p_2\to \p_{1'},\p_{2'})|^2$
is determined by the quasiparticle interaction vertex $\Gamma$ and
the quasiparticle weight $z$ \cite{AGD}.


\begin{figure}[t]
\includegraphics[width=1.9in,height=1.4in]{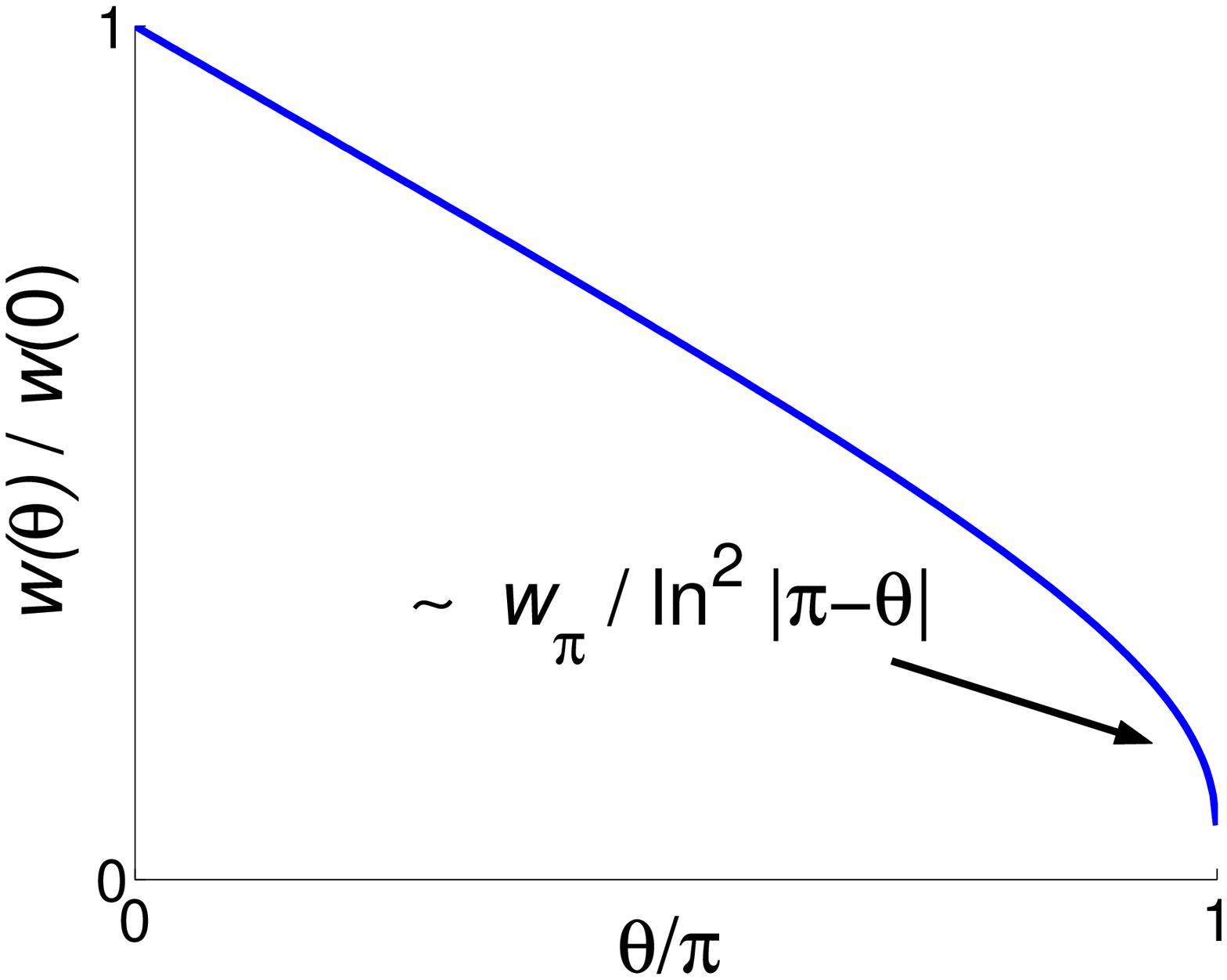}%
\includegraphics[width=1.5in]{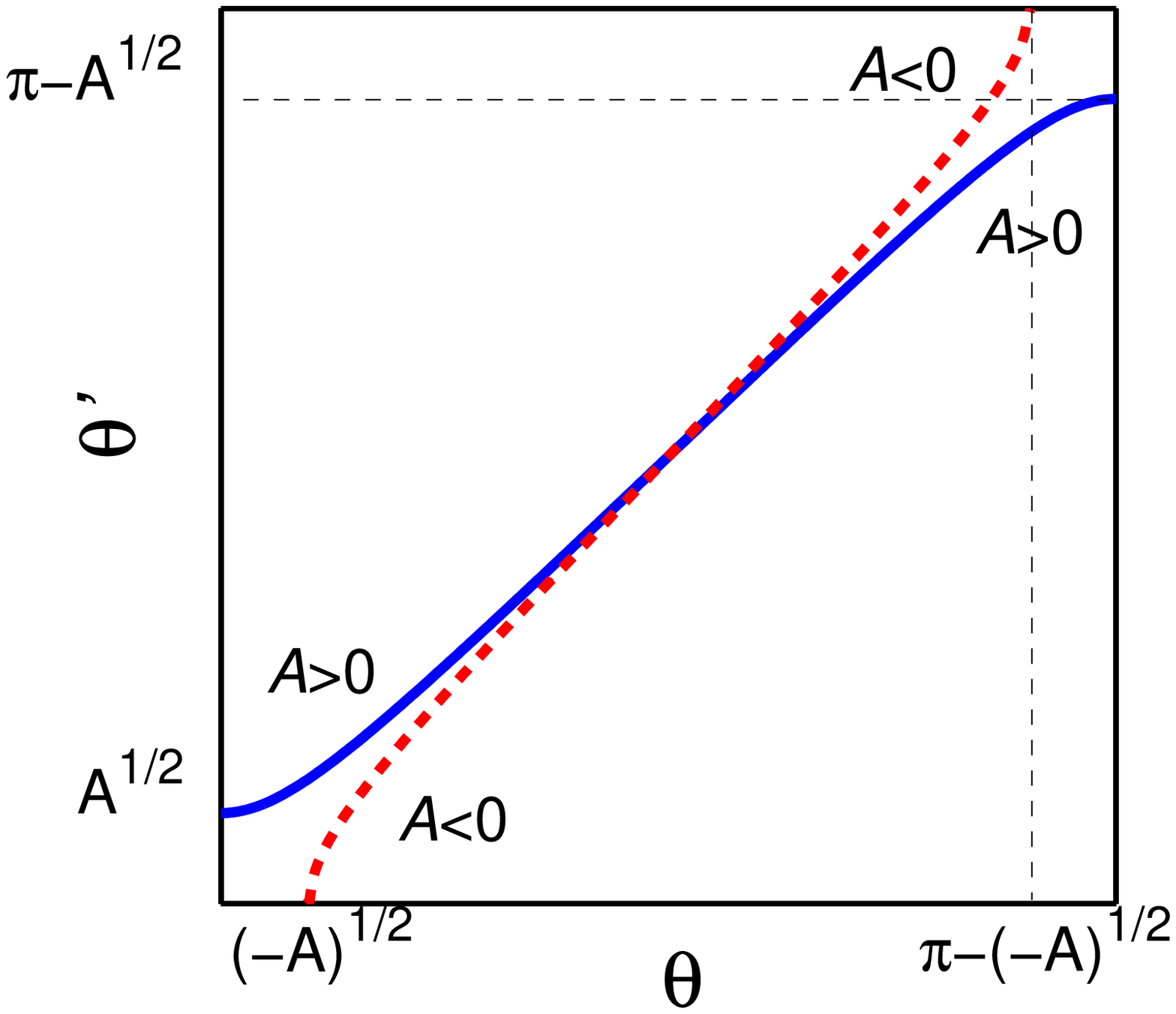}
\caption[]{
Left: Typical angular dependence of the quasiparticle scattering 
probability $w(\theta)$ that includes the $\ln^2|\pi-\theta|$
Cooper-channel singularity. 
Right: The relation between angles $\theta$ and $\theta'$ illustrated,
depending on ${\rm sign\,} (A)$, according to 
$\cos\theta' = (1-\frac{\bar A}2)\cos\theta$,
$|\bar A|\ll 1$.
}
\label{fig:w}
\end{figure}

We look for a solution $\psi$ of the linearized 
Eq.~(\ref{BE}) in the form of (\ref{LHS}), 
$\psi(\p) = - q(p) Y(\p,\u)$, where the angular part
\be \label{F}
Y(\p,\u) = 
\lp p_j {\partial \epsilon\over \partial p_i} - {\delta_{ij}\over 2} 
\p {\partial \epsilon\over\partial \p}\rp \cdot \frac12 
\lp \partial_i u_j + \partial_j u_i - \delta_{ij} \nabla \u \rp 
\ee
has the form of the second angular harmonic $Y\propto \cos 2(\phi_p - \phi_u)$,
$\phi_p = \arg (p_x + ip_y)$, 
with respect to the angle 
$\phi_u = \frac12 \tan^{-1} {\partial_x u_y + \partial_y u_x \over 
\partial_x u_x - \partial_y u_y}$ defined by the flow $\u(x,y)$. 
Substituting $\psi$ into the collision term (\ref{St-psi}),
we aim at the equation for the scalar deviation $q$ that has a meaning of the 
viscous relaxation time. 
The integration measure $d\tau_{1'}d\tau_2 
= \nu_F^2 d\xi_2 d\xi_{1'} (d\theta/2\pi) (d\theta_{1'}/2\pi)$,
where $\xi = \epsilon - \EF$, and $\nu_F=m^*/(\pi\hbar^2)$ is the 2d density 
of states.
The energy delta-function sets the value for the angle 
$\theta_{1'}=\angle (\p_{1'}, \p_1)$ via 
\be
\delta(\epsilon_1 + \epsilon_2 - \epsilon_{1'}-\epsilon_{2'})
= |\partial \epsilon_{2'}/\partial \theta_{1'}|^{-1}
\delta (\theta_{1'}-\theta_{1'}^{(0)})
\ee
with the Jacobian \cite{HSW}
\be \label{jacobian}
{\partial \epsilon_{2'}/ \partial \theta_{1'}} = 
-2 \lp \epsilon_1 \epsilon_2 \sin^2\theta + A\rp^{1/2} 
\simeq -2\EF |\sin\theta'|\,, 
\ee
where $\theta'=\angle(\p_{1'},\p_{2'})$ (Fig.~\ref{fig:collision}), and 
\be \label{A}
A \equiv \xi_{1'}\xi_{2'}- \xi_{1}\xi_{2}
= (\xi_1 - \xi_{1'})(\xi_{1'} - \xi_{2}) \,.
\ee
In the last part of Eq.~(\ref{jacobian}) 
we used $\cos\theta' \simeq (1-\frac{\bar A}2)\cos\theta$, $\bar A \equiv A/\EF^2$,
that follows from $\p_1\p_2=\p_{1'}\p_{2'}$. 
The relation between $\theta$ and $\theta'$ is illustrated in Fig.~\ref{fig:w}.

The symmetry of the integrand in Eq.~(\ref{St-psi}) 
with respect to $\theta \to -\theta$
allows us to substitute $Y_2 \to Y_1 \cos 2\theta$,  
$Y_{1'} \to Y_1 \cos 2\theta_{1'}$, and $Y_{2'} \to Y_1 \cos 2\theta_{2'}$,
and to cancel $Y_1\equiv Y(\p_1,\u)$ from the both sides of the 
Boltzmann equation.
Introducing the dimensionless energy variables $x=\xi/T$, 
as a result we arrive at the integral equation for $q(x)$:
\be
{f(-x_1)\over \nu_F^2 T^2} = 
\int \! 
\frac{f(x_2)f(-x_{1'})f(-x_{2'})w(\theta) Q d\theta dx_2 dx_{1'}}
{(2\pi)^2 \times 2\EF |\sin\theta'|}.
\label{BE-Q}
\ee
Here $x_{2'} = x_1+x_2-x_{1'}$, the integration in $\theta$ is between
$0$ and $\pi$ since the particles are indistinguishable, and 
\bea \nonumber
Q &\equiv& q(x_1) + q(x_{1'})(\cos2\theta - \cos 2\theta_{1'} - \cos 2\theta_{2'})
\\
&=& q_1- q_{1'} 
+ 2q_{1'}(\sin^2\theta_{1'} + \sin^2\theta_{2'} - \sin^2\theta) \,. \ 
\label{Q}
\eea
The stress tensor 
$\Pi_{ij} = \int \!d\tau \, \dn p_i {\partial \epsilon /\partial p_j}$,
combined with the definition (\ref{def-eta}), 
yields the viscosity in terms of $q(x)$: 
\be \label{eta-q}
\eta^{(2d)} = \frac14 N m^* v_F^2 \tau_\eta \,, \quad
\tau_\eta = \int\! {q(x) dx \over (2\cosh \frac{x}2)^2} \,,
\ee
where $N = \nu_F \EF$.
Above we used $\la \hat p_i \hat p_j \ra = \frac12 \delta_{ij}$ and 
$\la \hat p_i \hat p_j \hat p_k \hat p_l\ra = \frac18 
(\delta_{ij}\delta_{kl} + \delta_{ik}\delta_{jl} + \delta_{il}\delta_{jk})$.

{\it Quasiparticle lifetime.---}
The solution of the integral equation (\ref{BE-Q}) is related 
to the problem of the 2d quasiparticle lifetime $\tee$ 
that is determined via substituting $Q(x_1,x_{1'},x_2,\theta)$ by 
the isotropic term $\tee(x_1)$. Below we will focus on this simpler problem
first, to underscore the different roles played by the Cooper channel scattering
processes in the inelastic lifetime and in the viscosity.

To calculate $\tee$, one specifies the scattering probability $w(\theta)$
entering Eq.~(\ref{BE-Q}). In Refs.~\cite{HSW,Chaplik,Bloom,GQ,FA,ZDS,JMD,RW,NZA}
the scattering was assumed to be $\theta$-independent, $w={\rm const}$. 
In this case the forward scattering ($\theta=0$) and
the Cooper-channel ($\theta=\pi$) contribute equally to the logarithmic singularity
in $\tee$, via  
$\int_{\theta_0}^{\pi/2} \! 
{d\theta \over \sin\theta'} 
= \int_{\pi/2}^{\pi-\theta_0} \! 
{d\theta \over \sin\theta'} 
\simeq \ln (4|\bar A|^{-1/2})$,
$|\bar A|\ll 1$.
Here $\theta_0=0$ for $A>0$ and  $\sin\theta_0 = \sqrt{-\bar A}$ for $A<0$.

We now argue that the contributions of the forward scattering 
and of the backscattering are in fact parametrically different if 
one takes into account the Cooper ladder of diagrams
that are logarithmically divergent to all orders in the interaction.
Summing the ladder causes a singularity in 
$w(\theta\approx\pi;\theta_{1'})\simeq {2\pi\over\hbar}|z^2\Gamma(\p_1,\p_{1'}; {\bf s})|^2$, $\theta_{1'}=\angle(\p_1,\p_{1'})$, due to the
renormalization of the Cooper-channel interaction vertex 
\be \label{cooper}
\frac{\nu_F z^2}2 \Gamma(\p_1,\p_{1'}; {\bf s}) = 
\sum_{0,2,...} \tilde F_m^{s} \cos m\theta_{1'}
+\sum_{1,3,...} \tilde F_m^{t} \cos m\theta_{1'} 
\ee
(Ref.~\cite{AGD}, Sec. 20): In the leading-logarithm approximation,
$\tilde F_m = F_m^{(0)}/[1+\eta_m F_m^{(0)}\ln {p_F\over s}]$, 
$1/\eta_m=2-\delta_{m0}$, and $F_m^{(0)}$ are the bare couplings. 
[Recent RG treatment shows \cite{AleinerEfetov} that, in general,
even the asymptotic behavior $\tilde F_m= F_m/\ln(1/|\bar\theta|)$ 
is characterized by non-universal parameters $F_m \sim 1$.]
The above singularity suppresses the scattering
probability $w(\theta\sim \pi)\simeq w_\pi/\ln^2|\bar\theta|$,
where $\bar\theta=\pi-\theta \ll 1$ (Fig.~\ref{fig:w}, left panel),
and $w_\pi$ is dominated by the leading harmonic in Eq.~(\ref{cooper}).
Thus the lifetime $\tee$ is determined by the forward scattering,
$\int_{\theta_0}^{\pi-\theta_0} \! {w(\theta) d\theta \over \sin\theta'} 
\simeq  w(0) \ln\lp 4|\bar A|^{-1/2}\rp $, where we neglected
the backscattering contribution 
$\sim  \int_{\theta_0}^{\pi/2} \! 
{(F_0^{(0)s})^2 d\ln \bar\theta \over (1-F_0^{(0)s}\ln\bar\theta)^2}$.
The amplitude $w(0)$ is expressed in terms of the 
Landau parameters $\nu_F\f = F^\rho + F^\sigma {\vec \sigma}_1{\vec\sigma}_2$,
$F^{\rho,\sigma}(\theta)=\sum_{m=0}^\infty F_m^{\rho,\sigma}\cos m\theta$,
as 
\be \label{w0}
\nu_F^2 \overline{w(0)} \simeq {2\pi\over\hbar} F_0^2 \,, \quad 
F_0^2 \equiv \lp {F_0^\rho \over 1+F_0^\rho}\rp^2 
+ 3\lp {F_0^\sigma \over 1+F_0^\sigma}\rp^2 .
\ee
Here we approximated the probability $w(0)$ by its average over the Fermi surface
(neglecting the $\theta=\pi$ suppression) and averaged over the spin polarizations
\cite{footnote1}.
As a result \cite{footnote2},
\be \label{tau-qp}
{\hbar\over \tee(\xi,T)}  \simeq 
{\xi^2+\pi^2 T^2\over 8\pi\EF} F_0^2 \ln {\EF\over \sqrt{\xi^2+T^2}} \,.
\ee
The rate (\ref{tau-qp}) is twice smaller than that of 
Refs.~\cite{ZDS,JMD,NZA} as a consequence of 
the Cooper-channel renormalization \cite{footnote3}.


{\it The role of the Cooper channel in 2d.---}
The above calculation demonstrates that out of the two singularities
(at $\theta=0$ and $\pi$) of the integrand in Eq.~(\ref{BE-Q}), the former 
determines the lifetime $\tee$, while the latter is suppressed by the Cooper 
logarithm. Below we argue that in the case of the viscosity, it is the 
$\theta=\pi$ scattering that dominates, while other collision angles 
$\theta$, including $\theta=0$, are less relevant in $T/\EF$.
Indeed, for a generic 
$\theta$, 
the energy and momentum conservation select the forward scattering,
$\theta_{1'}\approx 0$ and $\theta_{2'}\approx \theta$ 
(see Fig.~\ref{fig:collision}). Due to the angular structure of Eq.~(\ref{Q}),
these contributions to the integrand are subdominant in the 
powers of $T/\EF$ since they
originate from the deviations of the colliding momenta from $p_F$.
This feature is specific to the 2d scattering.
In contrast, in 3d, the collision term for the viscosity is acquired from
all the scattering angles $\theta$ and $\phi$, 
where $\phi$ is the angle between the planes 
defined by the incoming and outgoing momenta. 
(The 3d collision term of 
Refs.~\cite{Abrikosov-Khalatnikov,Smith-Jensen-book} indeed becomes small
in $T/\EF$ for the in-plane scattering $\phi=0$ or $\pi$.)
For the head-to-head collisions ($\theta=\pi$),
the phase volume for scattering rapidly increases (similar to the situation
in BCS superconductivity), and there is no $T/\EF$ suppression of the integrand. 

{\it Viscosity calculation.---}
The main difference of the collision term (\ref{BE-Q}) from that 
for the inelastic lifetime is the complicated angular structure in Eq.~(\ref{Q}),
which we will focus on below. 
In accord with the above discussion, 
the main contribution comes from the region where
$\sin^2\theta_{1'}\sim 1$, in which case $|\bar\theta|, |\bar\theta'|\ll 1$, 
where $\bar\theta'=\pi-\theta'$ (right panel of Fig.~\ref{fig:collision}).
In this limit 
$\sin^2\theta_{1'}\approx \sin^2\theta_{2'} \gg \sin^2\theta\approx \bar\theta^2$,
such that $Q\approx q(x_1) - q(x_{1'}) + 4q(x_{1'}) \sin^2\theta_{1'}$. 
To estimate $\theta_{1'}$ in the limit $|\bar\theta|, |\bar\theta'|\ll 1$, 
we introduce the unit vector $\hat{\bf s}\simeq (p_1-p_2, p_F\bar\theta)/s$ in 
the direction of $\p_1+\p_2$ 
(here we chose the first coordinate axis parallel to $\p_1$),
and the unit vector $\hat{\bf s}' \simeq (p_{1'}-p_{2'}, p_F\bar\theta')/s$ that  
is rotated relative to ${\bf s}$ by $\theta_{1'}\approx\theta_{2'}$. 
Here $s^2=(p_1-p_2)^2+p_F^2\bar\theta^2=(p_{1'}-p_{2'})^2+p_F^2\bar\theta'^2$.
Then $\cos\theta_{1'} \simeq {\bf \hat s}{\bf \hat s}'$.
Since $\bar\theta'^2 \simeq \bar\theta^2 + \bar A$,
it is clear that for $\sin^2\theta_{1'}\sim 1$, one needs
either $\bar\theta^2\ll \bar\theta'^2$ for $A>0$, 
or $\bar\theta'^2 \ll \bar \theta^2$ for $A<0$. 
For such values of $\bar\theta$ and $\bar\theta'$, 
$s/p_F\simeq {\rm max\ }\{ |\bar\theta|,|\bar\theta'| \}$, and we estimate
\be \label{cos-theta1'}
\cos^2\theta_{1'} \simeq {\rm min\ } 
\lf \lp {\bar\theta/\bar\theta'}\rp^2 , \
\lp {\bar\theta'/ \bar\theta}\rp^2 \rf . 
\ee
[Dropping the $(p_1-p_2)(p_{1'}-p_{2'})$ terms in 
${\bf \hat s}{\bf \hat s}'$ is justified since their contribution will be 
smaller in powers of $T/\EF$ after the energy integration.]
%
%
To perform the angular integration in Eq.~(\ref{BE-Q}) we 
keep the lowest singlet and triplet harmonics in the series (\ref{cooper}),
and assume $\tilde F_{0}^s \simeq F_{0}^s/\ln(1/|\bar\theta|)$ 
and $\tilde F_{1}^t \simeq F_{1}^t/\ln(1/|\bar\theta|)$ \cite{AleinerEfetov}. 
Then their contributions are
\bea \label{angular-singlet}
\int_0^{\bar\theta'}\! {(F_0^s)^2 d\bar\theta \sin^2\theta_{1'} 
\over \ln^2|\bar\theta| \sqrt{\bar A + \sin^2\bar\theta}}
\simeq 
{(F_0^s)^2\over \ln^2\sqrt{\bar A}} 
\left. {\bar\theta\over \bar\theta'}\right|_{\bar\theta =0}^{\bar\theta =\bar\theta'}
= {(F_0^s)^2\over \ln^2\sqrt{\bar A}} , & \quad \\
\label{angular-triplet}
\int_0^{\bar\theta'}\! {(F_1^t)^2 \cos^2\theta_{1'}
d\bar\theta \sin^2\theta_{1'}
\over  \ln^2|\bar\theta| \sqrt{\bar A + \sin^2\bar\theta}}
\simeq   
{(F_1^t)^2\over 3\ln^2\sqrt{\bar A}} &
\eea
(here $A>0$). For $A<0$, the calculation is similar by interchanging
$\bar\theta \leftrightarrow \bar\theta'$, 
$\bar\theta d\bar\theta=\bar\theta' d\bar\theta'$, yielding the above results
with $\bar A \to |\bar A|$ under the logarithms.
Thus the net angular contribution is $\frac14 F_\pi^2/\ln^2\sqrt{|\bar A|}$,
where we introduced the spin-averaged dimensionless coupling $\frac14 F_\pi^2$, 
with $F_\pi^2 \equiv (F_0^s)^2  + (F_1^t)^2$. 
We plug Eqs.~(\ref{angular-singlet}) and (\ref{angular-triplet}) 
into Eq.~(\ref{BE-Q}), use the standard integrals
$
\int \! dx_2 dx_{2'} \, 
f(x_2)f(-x_{1'})f(-x_{2'}) 
= f(-x_{1}) {\cosh(x_1/2)\over \cosh(x_{1'}/2)} \K(x_1-x_{1'})$,
where $\K(x) = x / (2 \sinh \frac{x}2)$, and 
$ 
%
\int \! dx_{1'} dx_2 dx_{2'} \, 
f(x_2)f(-x_{1'})f(-x_{2'}) 
=\frac12 f(-x_{1}) (\pi^2 + x_1^2)
$,
and obtain the integral equation 
\be \label{IE}
{1\over \cosh {x_1\over 2}} = (\pi^2 + x_1^2)y(x_1) - \int\! dx_{1'} 
\tilde \K(x_1-x_{1'}) y(x_{1'}) 
\ee
for the function
$y(x_1) \equiv {q(x_1) \over 2 \tau_0 \cosh (x_1/2)} 
\ln {\EF \over T\sqrt{1+x_1^2}} 
$,
where 
\be \label{tilde-K}
\tilde \K(x) = \K(x)\times 
2\lb 1 -  \lp{F_\pi \over F_0}\rp^2  \ln^{-3} {\EF\over T\sqrt{1+x^2}} \rb ,
\ee
and 
${\hbar/ \tau_0} \equiv { T^2/(4\pi \EF)}$ \cite{footnote2}.


To solve Eq.~(\ref{IE}), we note that the viscosity (\ref{eta-q})
is determined by $\xi\lesssim T$, 
which allows us to set $x=0$ under the logarithms,
$\ln 1/\sqrt{|\bar A|} \simeq \ln (\EF/T\sqrt{1+x^2})\approx \ln(\EF/T)$,
reducing it to the standard problem \cite{Brooker-Sykes,Smith-Jensen-book},
whose solution yields
\be
\tau_\eta = {\tau_0 \over 2 \ln{\EF\over T}} 
\lf \frac13 + \frac{4\alpha}{\pi^2}\sum_{1,3,...} {2n+1\over n^2(n+1)^2}
{1\over n(n+1)-\alpha} \rf ,
\ee
where $\alpha = 2\lb 1-(F_\pi/F_0)^2/\ln^3(\EF/T)\rb$.
Since $2-\alpha \ll 1$, from Eq.~(\ref{eta-q}) we obtain the result (\ref{eta-2d}).
As advertized, the viscosity is fully determined by the quasiparticle 
scattering probability 
$\nu_F^2 \overline{w}\simeq {2\pi\over\hbar} \frac14 F_\pi^2/\ln^2(\EF/T)$ 
in the Cooper channel.

In summary, we found a nonanalytic temperature behavior of the 2d Fermi liquid
viscosity and related it to the logarithmic singularity of the 
quasiparticle scattering amplitude in the Cooper channel. 
The ratio between the viscous and the inelastic scattering times is 
enhanced by the factor $\sim \ln^3(\EF/T)$ due to the restrictions on the 
2d quasiparticle scattering.

It is a pleasure to thank I. Aleiner and B. Altshuler for helpful discussions.
This work was supported by NSF MRSEC grant DMR 02-13706.



\begin{thebibliography}{99}
\bibitem{footnote2}
The argument of $\ln$ is chosen to interpolate between the limits
$\xi\ll T$ and $\xi\gg T$ within the adopted accuracy.
\bibitem{footnote3}
In Ref.~\cite{ML} this renormalization was considered, but the final 
answer has an extra factor of $\frac12$ caused by an algebraic error.
\end{thebibliography}
\end{document}